\documentclass[11pt]{article}
\usepackage{amsmath,color}
\usepackage{amssymb}
\usepackage{amsfonts}
\usepackage{graphicx}
\usepackage{color}

\definecolor{darkgreen}{rgb}{0,0.35,0}

\numberwithin{equation}{section}

\begin{document}

\title{\textbf{Evanescent gravitons in Warped Anti-de Sitter space}}
\author{Gaston Giribet$^{1,2,3}$, \ Yerko V\'asquez$^4$}
\maketitle

\begin{center}

\smallskip
\smallskip
\centerline{$^1$ Universit\'{e} Libre de Bruxelles and International Solvay Institutes}
\centerline{{\it ULB-Campus Plaine CPO231, B-1050 Brussels, Belgium.}}

\medskip
\textit{$^{2}$}Departamento de F\'{\i}sica, Universidad de Buenos Aires
FCEN-UBA and IFIBA-CONICET,

\textit{Ciudad Universitaria, Pabell\'{o}n I, 1428, Buenos Aires, Argentina.}

\medskip
\textit{$^{3}$}Instituto de F\'{\i}sica, Pontificia Universidad Cat\'olica
de Valpara\'{\i}so,\textit{\ }

\textit{Casilla 4950, Valpara\'{\i}so, Chile.} \\[0pt]

\medskip
\textit{$^{4}$}Departamento de F\'{\i}sica, Universidad de la Serena,\textit{%
\ }

\textit{Avenida Cisternas 1200, La Serena, Chile.} \\[0pt]

\end{center}

\bigskip

\bigskip

\bigskip

\bigskip

\begin{abstract}

Besides black holes, the phase space of three-dimensional massive gravity about Warped Anti-de Sitter (WAdS) space contains solutions that decay exponentially in time. They describe evanescent graviton configurations that, while governed by a wave equation with non-vanishing effective mass, do not carry net gravitational energy. Explicit examples of such solutions have been found in the case of Topologically Massive Gravity; here, we generalize them to a much more general ghost-free massive deformation, with the difference being that the decay rate gets corrected due to the presence of higher-order terms.  

\end{abstract}

\newpage

\section{Introduction}

It has been shown in \cite{Proceeding} that the holographic description of the entropy of Warped Anti-de Sitter (WAdS) black holes in three-dimensional massive gravity still holds when one considers the asymptotic boundary conditions proposed in Ref. \cite{Enoc}, which allow for new asymptotically WAdS$_3$ solutions that have exactly the same conserved charges that the black holes. In other words, the new geometries seem not to contribute substantially to the ensemble, and the holographic computation based on the asymptotic charge algebra \cite{DHH, nos2} exactly reproduces the entropy of the WAdS$_3$ black holes even in the case of the new boundary conditions. Here, with the aim of understanding the extent of this phenomenon, we further study the configuration space of three-dimensional massive gravity in asymptotically WAdS$_3$ space. We construct exact solutions that represent time-dependent deformations of the WAdS$_3$ black holes \cite{BHs} and preserve no isometries. The solutions decay exponentially in time and represent evanescent gravitons that, while obeying a wave equation with non-vanishing effective mass, do not carry net gravitational energy. Such geometries had previously been observed to exist in Topologically Massive Gravity (TMG) \cite{TMG}; here, we extend the set to the model that, in addition of the gravitational Chern-Simons term, includes both the higher-derivative terms of the so-called New Massive Gravity \cite{NMG} and the higher-curvature terms of the recently proposed Minimal Gravity \cite{MMG}. We show that the decay rate receives corrections due to the presence of higher-order terms, which means that the evanescent solutions, unlike the black holes, are not locally equivalent to empty WAdS$_3$ space.

\section{Massive gravity}

The field equations of the general massive gravity we will consider are
\begin{equation}
 G_{\mu \nu}+\Lambda g_{\mu \nu }+\gamma J_{\mu \nu}+\frac{1}{\mu}C_{\mu \nu}-\frac{1}{2m^2}K_{\mu \nu}=0, \label{uno}
\end{equation}
where $G_{\mu \nu}$ is the Einstein tensor; the tensor $J_{\mu \nu }$ is the contribution of the Minimal Massive Gravity (MMG) \cite{MMG}, which can be written as
\begin{equation}
 J_{\mu \nu}=\frac{1}{2} \frac{\epsilon_{\mu}^{\,\,\,\, \rho \sigma}}{\sqrt{-g}} \frac{\epsilon_{\nu}^{\,\,\,\, \tau \eta}}{\sqrt{-g}} S_{\rho \tau} S_{\sigma \eta}, \ \ \ \ \ \ \text{with} \ \ \ \ \ S_{\mu \nu}= R_{\mu \nu}-\frac{1}{4} g_{\mu \nu}R ; \label{Jota}
\end{equation}
$C_{\mu \nu }$ is the Cotton tensor of Topologically Massive Gravity (TMG) \cite{TMG}, namely
\begin{equation}
 C_{\mu \nu}=\frac{\epsilon_{\mu}^{\,\,\,\, \tau \rho}}{\sqrt{-g}} \nabla_{\tau} S_{\rho \nu}; \label{Ce}
\end{equation}
and $K_{\mu \nu }$ is the contribution of New Massive Gravity (NMG) \cite{NMG}, namely
\begin{equation}
K_{\mu \nu}=2 \Box S_{\mu \nu}-\frac{1}{2} \nabla_{\mu} \nabla_{\nu} R  +4 R^{\alpha \beta} S_{\mu \alpha \nu \beta} -\frac{3}{2} R S_{\mu \nu}, \label{K}
\end{equation}
with $S_{\mu \alpha \nu \beta}\equiv R_{\mu \alpha \nu \beta} -({1}/{4}) g_{\mu \nu } R_{\alpha \beta }$. In three dimensions, the Riemann tensor can be expressed in terms of the Ricci tensor; nevertheless, it is convenient to write the expression of $K_{\mu \nu }$ as in (\ref{K}).

Being of second order in the metric, tensor (\ref{Jota}) can not be derived from an action in the metric formalism. This can also be concluded from the fact that $J_{\mu \nu }$ is not covariantly conserved identically; although it is covariantly conserved on-shell \cite{MMG}. In contrast, tensors (\ref{Ce}) and (\ref{K}) do follow from an action principle \cite{TMG, NMG}. The former tensor is conformally invariant, satisfying $C_{\mu}^{\ \mu}=0$, while the latter is conformally covariant, in the sense that its trace, $K_{\mu}^{\ \mu}$, coincides with the Lagrangian from which it is derived. Despite the fact that $C_{\mu \nu }$ and $K_{\mu \nu }$ are of third and fourth order in the metric, respectively, the trace of the field equations (\ref{uno}) remains of second order; in fact, one finds $C_{\mu }^{\ \mu}=K_{\mu }^{\ \mu}+2J_{\mu }^{\ \mu}=0$. This property is one of the reasons why the theory (\ref{uno}) turns out to be free of ghosts \cite{NMG} around suitable backgrounds. On the other hand, the inclusion of the term $J_{\mu\nu}$ in the field equations permits to cure the so-called bulk/boundary unitarity clash that the theory (\ref{uno}) with $\gamma =0$ suffers from when formulated around AdS$_3$ \cite{MMG} (see also \cite{Kuko7}-\cite{Kuko8}). Therefore, all the terms of (\ref{uno}) being important to define a well-behaved theory of massive gravity in three dimensions, we find appropriate to consider all of them when studying the theory formulated around Warped deformations of AdS$_3$ space. 



\section{Warped black holes}

We will be concerned with geometries that asymptote to WAdS$_3$ space. The latter space corresponds to a one-parameter stretched deformation of AdS$_3$. Our conventions will parallel those of Ref. \cite{Aninnos}. In such language, the metric of the asymptotically WAdS$_3$ black hole solutions \cite{BHs} reads
\begin{eqnarray}
\notag d{s}_{\text{BH}}^2 &=& dt^2+\frac{\ell^2}{\nu^2+3}\frac{dr^2}{(r-r_{+})(r-r_{-})}-2 \left( \nu r+\frac{1}{2}\sqrt{r_{+}r_{-} (\nu^2+3)}\right)dt d\phi \\
&& +\frac{r}{4}\left( 3(\nu^2-1) r+(\nu^2+3) (r_{+}+r_{-})+4\nu \sqrt{r_{+}r_{-} (\nu^2+3)} \right) d\phi^2, \label{warped}
\end{eqnarray}
where $\nu $ is the parameter that controls the stretching deformation (with $\nu^2 \geq 1$ and with the case $\nu =1$ corresponding to unstretched AdS$_3$) and $\ell $ is the curvature radius. Metrics (\ref{warped}) for $r_+\geq r_-\geq 0$ represent black hole solutions that asymptote to WAdS$_3$ space at large $r$. These solutions exhibit two horizons, located at $r=r_+$ and $r=r_-$. Being locally equivalent to WAdS$_3$ space, the black hole solutions (\ref{warped}) solve the field equations for the same values of parameters than the former; more precisely, they are solutions of (\ref{uno}) provided the following two conditions are satisfied
\begin{eqnarray}\label{relations}
24 m^2 \ell^2 \nu (\nu^2-1)+\ell \mu \left(9-84\nu^2+76\nu^4+m^2 (4 \ell^4 \Lambda +\ell^2 (12-8 \nu^2)+\gamma (3-2 \nu^2)^2) \right)=0,
\end{eqnarray}
and
\begin{eqnarray}\label{relations2}
3(\nu^2-1) \left(9 \ell \mu \nu^2 (2 \nu^2-1)+m^2 (\ell^3 \mu (3+2 \ell^2 \Lambda -2\nu^2)+3 \ell^2 \nu (2\nu^2-1))  \right)=0.
\end{eqnarray}

As expected, for $\nu =1$, where the stretching effect disappears, the latter equation gives no restriction for the coupling constants. For $\nu =1$, and for fixed $\mu $, in the limit $m^2\to \infty $ and with $\gamma = 0$, equation (\ref{relations}) yields the expected result $\Lambda = -\ell^{-2}$. On the other hand, for arbitrary $\nu $, in the limit $\mu\to\infty$ and with $\gamma =0$, one recovers the result for NMG \cite{Clement}, $
m^2\ell^2=({20\nu^2-3})/{2}$.

\section{Evanescent gravitons}

Now, let us construct asymptotically WAdS$_3$ time-dependent geometries. The procedure we will employ to construct such solutions is similar to that of Refs. \cite{Eloy, AdSwaves, GAY}, here adapted to the case of WAdS$_3$. It amounts to consider a Kerr-Schild ansatz that deforms the locally WAdS$_3$ geometries (\ref{warped}). To see how it works, and before going to the general case, let us begin by considering a deformation of the zero mass black hole. The latter is given by setting $r_{+}=r_{-}=0$ in (\ref{warped}); namely
\begin{equation}
ds^2_{0}=dt^2+\frac{\ell^2}{\nu^2+3} \frac{dr^2}{r^2}-2\nu r dt d\phi +\frac{3}{4} (\nu^2-1)r^2 d\phi^2 . \label{lacero}
\end{equation}

Then, consider a deformation of the form 
\begin{equation}\label{metric1}
ds^2=d{s}_{0}^2+H(t,r) k_{\mu} k_{\nu} dx^{\mu} dx^{\nu},
\end{equation}
with $k_{\mu }$ being a null vector. It is convenient to chose
\begin{equation}
k_{\mu}dx^{\mu }\equiv \ell dv, \ \ \ \ \text{with} \ \ \ \ v=\frac{2\ell }{(\nu ^2+3)r}+\phi .
\end{equation}

$H(t,r)$ is a function that will be demanded to be eigenfunction of the D'Alembert operator associated to (\ref{lacero}); that is, $\Box H(t,r)=M^2H(t,r)$ with $M^2\in \mathbb{R}$. To solve the latter equation, the form of the operator suggests to propose the ansatz $H(t,r)\propto e^{-\omega t} r^{\delta } $. By plugging it back into the field equations (\ref{uno}), we find two conditions between $\omega $, $\delta $, and the coupling constants of the theory. These conditions read
\begin{eqnarray}
P(\nu , \omega ) &\equiv &  m^2 \ell^2 (-20\nu^6 -24\nu^4+63\nu^2-27-2\ell^2 \Lambda (26 \nu^4-27 \nu^2+9) \nonumber \\
&&-6 \omega \ell  \nu (4\nu^4+2(\ell^2 \Lambda -3) \nu^2+3) +\omega^2\ell^2 (-2\nu^2+2 \ell^2 \Lambda +3)(2 \nu^2-3) ) \nonumber \\
&&+3 \nu (4 \nu^4-8 \nu^2+3) (\nu+\ell \omega )((5\nu+\ell \omega)(\nu+\ell \omega)+3)=0, \label{polinomia}
\end{eqnarray}
and 
\begin{equation}
\delta = -\frac{2\nu\ell\omega }{(\nu^2+3)}. 
\end{equation}

After having determined the possible values for $\delta $ and $\omega$, given by the real roots of (\ref{polinomia}), one can verify that the perturbed metric (\ref{lacero}) remains a solution if one multiplies $H(t,r)$ by an arbitrary function of $v$. 

The same procedure can be applied to the black hole metric (\ref{warped}) with $r\neq r_{\pm }$. Considering a generalization of the above metric, namely
\begin{equation}
ds^2=d{s}_{\text{BH}}^2+H(t,r,\phi) k_{\mu} k_{\nu} dx^{\mu} dx^{\nu}, \label{sol1}
\end{equation}
with the null vector
\begin{equation}
k_{\mu} dx^{\mu}= -\ell du , \ \ \ \ \text{with} \ \ \ \ u \equiv \frac{2\ell }{(\nu^2+3)(r_+-r_-)} \log \left( \frac{r-r_-}{r-r_+}\right) -\phi ,
\end{equation}
one finds that a solution of the field equations (\ref{uno}) is given by
\begin{equation}
H(t,r,\phi)=e^{-\omega t} \left( \frac{(r-r_{-})^{2 \nu r_{-}+\sqrt{r_{+}r_{-} (\nu^2+3)}}}{(r-r_{+})^{2 \nu r_{+}+\sqrt{r_{+}r_{-} (\nu^2+3)}}} \right)^{\frac{\omega \ell}{(r_{+}-r_{-}) (\nu^2+3)}} F(u ), \label{sol2}
\end{equation}
where $\omega $ is again given by (\ref{polinomia}), and where $F(u)$ is an arbitrary function of $u$; periodicity of $\phi $ demands, however, that $F(u)=F(u+2\pi )$. The arising of an arbitrary function of the null coordinate $u$ is usual in this type of solutions, and similar to what happens with $pp$-waves and AdS-wave solutions. Actually, solution (\ref{sol1}) can be thought of as a WAdS-wave deformation of the black hole background (\ref{warped}), with function $F(u)$ characterizing the profile of the wave. This analogy becomes even more clear when one finds that the profile function actually satisfies a wave equation (see (\ref{lamasa}) below).

For $\delta <0$, namely $\omega \ell \nu >0$, the solutions (\ref{sol1})-(\ref{sol2}) behave asymptotically as the solutions of TMG found in Refs. \cite{Enoc}, although with a different falling-off coefficient $\delta $. In this sense, they correspond to asymptotically WAdS$_3$ spaces, cf. \cite{Compere3}. It has been pointed out in \cite{Proceeding} that the boundary conditions of \cite{Enoc} are still compatible with the WAdS/CFT correspondence \cite{Aninnos}, at least in what regards the description of black hole thermodynamics \cite{DHH, nos2}.

At the horizon, for $r$ close to $r_+$, solutions (\ref{sol1})-(\ref{sol2}) behave as
\begin{equation}
H(t,r,\phi) \sim (r-r_+)^{- \frac{2\nu\omega\ell }{(r_+-r_-)(\nu^2+3)}( r_++\frac{1}{2\nu}\sqrt{(\nu^2+3)r_-r_+})} e^{-\omega t},
\end{equation}
while at the boundary, at large $r$, they behave as
\begin{equation}
H(t,r,\phi) \sim r^{-\frac{2\nu\omega\ell }{(\nu^2+3)}} e^{-\omega t}.
\end{equation}

This means that for asymptotically WAdS$_3$ solutions, namely for $\delta <0$, the metric function $H(t,r,\phi ) $ diverges at the horizon. In this sense, these solutions are reminiscent of those of Refs. \cite{GAY}. Nevertheless, all the curvature invariants remain finite at $r=r_{\pm }$; in fact, the curvature invariants are constant and independent of $F(u)$. When evolving in time, the region $r >> r_+$  of a solution with $\omega >0$ rapidly tends to the black hole geometry. Close to $r_+$, in contrast, such a solution develops a {\it wall}, characterized by divergence of $H(t,r,\phi )$ at $r=r_+$, which becomes thinner and thinner with time. Provided $\delta <0$, the solutions tend to WAdS$_3$ space at large $r$, at any $t$. That is to say, they respect the asymptotic isometries of WAdS$_3$, which are generated by a copy of the Witt algebra in semi-direct sum with a loop algebra of $u(1)$ \cite{Enoc} (we will come back to this point latter). However, the solutions with $\omega \neq 0$ and $\partial_{u}F(u)\neq 0$ do not preserve any exact isometry. The profile of such a solution resembles a helicoid in the $t=const$ plane, whose amplitude decreases as $r^{\delta }$ for $r$ sufficiently large. This means that solutions (\ref{sol1})-(\ref{sol2}) represent evanescent gravitons.

Solution (\ref{sol1})-(\ref{sol2}) generalizes to the theory (\ref{uno}) the solutions found in \cite{Enoc, nos2} for the particular cases of TMG and NMG. In fact, notice that by taking the limit $m^2\to \infty $ in (\ref{polinomia}) one actually recovers the result for TMG \cite{Enoc}, namely
\begin{equation}
P_{\text{TMG}}(\nu , \omega )= (5\nu +\omega  \ell) (\nu + \omega \ell) +3=0, \label{polinomio2}
\end{equation}
while in the limit $\mu \to \infty$ and with $\gamma =0 $, one recovers the result of NMG \cite{nos2}
\begin{equation}
P_{\text{NMG}}(\nu , \omega )= -4\nu\omega^2\ell^2-6\omega\ell +10\omega\ell \nu^2+16\nu^3-\omega^3\ell^3 =0. \label{polinomio3}
\end{equation}

The general polynomial equation (\ref{polinomia}) that determines $\omega $ as a function of $\nu $, also depends on the coupling constant $m^2$ and, through $\Lambda $, on $\mu $ and $\gamma $. This means that the decay rate of the time-dependent solution gets corrected with respect to that of TMG due to the presence of the ${\mathcal O}(R^2)$ terms; cf. Ref. \cite{Enoc}. This also implies that the solutions (\ref{sol1})-(\ref{sol2}), unlike the WAdS$_3$ black holes, are not locally equivalent to WAdS$_3$ space; we will discuss this latter in more detail.

\section{Gravitational energy}

The effective mass $M$ associated to $H(t,r,\phi )$ is given in terms of $\omega $ by $M^2=\omega ( \omega-{2 \nu}/{ \ell})$. That is, the function that describes the profile of the deformation is governed by the wave equation
\begin{equation}
\Box H(t,r,\phi)= \omega \left( \omega-\frac{2 \nu}{ \ell} \right)H(t,r,\phi) .  \label{lamasa}
\end{equation}

This is similar to what happens with the massive gravitational waves in AdS$_3$ space \cite{Eloy, AdSwaves}. It is worthwhile discussing the critical value $\ell \omega = 2\nu $, for which the effective mass $M$ vanishes. This corresponds to $\delta = -4\nu^2/(\nu^2+3)$. Notice that this value can not occur neither for TMG nor for NMG; while for the former this value corresponds to $\nu^2=1/7$, for the latter it corresponds to $\nu^2=3/5$. In both cases, this does not obey the constraint $\nu^2>1$ for the solution to represent a black hole and not to exhibit closed timelike curves.  

Although the deformation (\ref{sol2}) obeys a wave equation with non-vanishing effective mass, it turns out that the evanescent solutions do not carry net gravitational energy. This is analogous to what happens with electromagnetic evanescent waves in conducting media. To see explicitly that the evanescent gravitons do not carry gravitational energy, we can compute the mass associated to solutions (\ref{sol1})-(\ref{sol2}). In the cases of TMG and NMG, this can be done by standard methods \cite{ADT1}-\cite{IyerWald}. The case of MMG, however, is a little bit more subtle because of the property of $J_{\mu\nu }$ of not being conserved off-shell. In the case of asymptotically AdS$_3$ spaces, the charges of the latter theory have been constructed in \cite{Tekin:2014jna}, and in principle the definition therein can be extended to warped deformations. Here, we will consider the examples of TMG and NMG: Consider first the case $\gamma = 1/m^2 = 0 $ in (\ref{uno}). If one denotes by $g_{\mu \nu }$ the spacetime metric and by $\bar{g}_{\mu \nu }\equiv g_{\mu \nu }-h_{\mu \nu }$, the background metric respect to which the charges will be computed, then the charges are given by the following formula
\begin{equation}
Q\left( \bar{\xi}\right) =\frac{1}{16\pi G} \int_{\partial \mathcal{M}} \sqrt{-\bar{g}}\left(\mathcal{F}_{E}^{0 i}(\bar{\xi})
+
\frac{1}{\mu }\mathcal{F}_{C}^{0 i}(\bar{\xi})\right) dS_{i} ,  \label{Q}
\end{equation}
where $\bar{\xi}$ generates the asymptotic Killing symmetries that both $g_{\mu \nu }$ and $ \bar{g}_{\mu \nu }$ preserve. The integral in (\ref{Q}) is evaluated on the circle at spatial infinity. The functions 
$\mathcal{F}_{E}^{\mu \nu}(\bar{\xi})$ and $\mathcal{F}_{C}^{\mu \nu}(\bar{\xi})$ are defined as in \cite{Clemento}; they read
\begin{eqnarray}
\mathcal{F}_{E}^{\mu \nu}(\bar{\xi}) &=&\frac{1}{2}(\bar{\xi ^{\nu }}\bar{\nabla}%
_{\lambda }h^{\lambda \mu }-\bar{\xi ^{\mu }}\bar{\nabla}_{\lambda}h^{\lambda \nu }+\bar{\xi
_{\lambda }}\bar{\nabla}^{\mu}h^{\lambda \nu}-\bar{\xi _{\lambda}}\bar{\nabla}^{\nu }h^{\lambda \mu}+\bar{\xi}^{\mu}\bar{\nabla}%
^{\nu }h- \notag \\
&&\bar{\xi}^{\nu }\bar{\nabla}^{\mu }h+h^{\nu \lambda }%
\bar{\nabla}_{\lambda}\bar{\xi}^{\mu }-h^{\mu \lambda }\bar{\nabla}_{\lambda }\bar{\xi}^{\nu }+h\bar{\nabla%
}^{ [\mu }\bar{\xi}^{\nu ]}), \\
\mathcal{F}_{C}^{\mu \nu}(\bar{\xi}) &=&\mathcal{F}_{E}^{\mu \nu}(\bar{\Xi})+\frac{1}{\sqrt{-\bar{g}}} \bar{\xi}_{\lambda} \left( \epsilon ^{\mu \nu \rho} G^{L \lambda }_{\rho}-\frac{1}{2} \epsilon^{\mu \nu \lambda} 
G^{L \rho }_{\rho } \right) \notag \\
&+&\frac{1}{2\sqrt{-\bar{g}}} \epsilon^{\mu \nu \rho} \left( \bar{\xi}_{\rho} h^{\lambda}_{\sigma} \bar{G}^{\sigma}_{\lambda}+\frac{1}{2}h \left( \bar{\xi}_{\sigma} \bar{G}^{\sigma}_{\rho} +\frac{1}{2} \bar{\xi}_{\rho} \bar{R} \right) \right),
\end{eqnarray}%
where $\bar{\Xi}^{\beta }=\epsilon ^{\alpha \nu \beta }\bar{\nabla}_{\alpha }\bar{\xi}_{\nu }/(2\sqrt{-\bar{g}})$. $G^{L}_{\mu \nu }$ is the
linearized Einstein tensor, with the linearized Ricci tensor given by 
\begin{equation}
R_{\mu \nu }^{L} =\frac{1}{2}(-\bar{\Box}h_{\mu \nu }-\bar{\nabla}_{\mu
}\bar{\nabla}_{\nu }h+\bar{\nabla}^{\sigma }\bar{\nabla}_{\nu }h_{\sigma \mu
}+\bar{\nabla}^{\sigma }\bar{\nabla}_{\mu }h_{\sigma \nu }).
\end{equation}

For the timelike Killing vector $\bar{\xi ^{\mu }}=-\partial _{t}$, charge (\ref{Q})
corresponds to the gravitational energy. Choosing as the background metric $ds_0^2=\bar{g}_{\mu \nu } dx^{\mu}dx^{\nu}$ the
black hole with $r_+=r_-=0$, the gravitational energy of the configuration (\ref{sol1})-(\ref{sol2}) results to be 
\begin{equation}
E = \frac{(\nu^2+3)}{24 \ell G}  \left(r_{+}+r_{-}+\frac{1}{\nu} \sqrt{r_{+} r_{-}(\nu^2 +3)} \right),
\end{equation}
and is independent of the function $F(u)$. This is in agreement with the result found in \cite{Enoc}. In addition, same phenomenon can be shown to hold when the contribution of the higher-derivative terms of NMG, $\Delta E$, is included. To see this explicitly, one can resort to the result of Ref. \cite{Corea}, where the canonical method \cite{IyerWald, Barnich:2001jy, Barnich:2007bf} was implemented for the case of NMG. The general result for both $1/m^2\neq 0$ and $ 1/\mu \neq 0$ reads
\begin{equation}
E+\Delta E = \frac{(\nu^2+3)(4\nu^3 -8\ell\mu \nu^2 +3\nu)}{8 \ell^2 G \mu (3-20\nu^2)}  \left(r_{+}+r_{-}+\frac{1}{\nu} \sqrt{r_{+} r_{-}(\nu^2 +3)} \right), \label{masatotal}
\end{equation}
where the mass parameter is $m^2=\mu (20\nu^2-3)/(2\ell^2\mu -6\ell\nu )$. 

In conclusion, the mass of the solution is actually provided by the black hole geometry and the evanescent gravitons do not carry net gravitational energy. However, this does not mean that the time-dependent gravitons behave as stealth probes that do not backreact on the geometry. On the contrary, the deformation (\ref{sol2}) does interact and it affects the fields in a non-trivial way. A simple toy model to illustrate this is provided by the scalar matter theory defined by the action
\begin{equation}
S=\int d^3x \sqrt{-g} \left( \zeta R -2\Lambda +\eta g^{\mu \nu } \nabla_{\mu}\varphi \nabla_{\nu}\varphi +\beta G^{\mu \nu } \nabla_{\mu}\varphi \nabla_{\nu}\varphi \right),
\end{equation}
which, as shown in \cite{Minas}, admits WAdS$_3$ black holes (\ref{warped}) as exact solutions provided the conditions $
\beta \nu^2=\eta \ell^2$ and $\zeta \nu^2=\Lambda \ell^2$ are satisfied. The scalar field configuration supporting the black hole geometry is given by
\begin{equation}
\varphi(r) = \sqrt{\frac{2\zeta \nu^2}{\eta (\nu^2+3)}} \log (\sqrt{r-r_+} +\sqrt{r-r_-}) +\varphi_0 , \label{FI}
\end{equation}
where $\varphi_0$ is an arbitrary constant. It can be easily shown that turning on the time-dependent deformation (\ref{sol2}) affects the background field (\ref{FI}) drastically and no static configuration exist for $\omega \neq 0$.

\section{Comments on WAdS/CFT}

The existence of solutions (\ref{sol2}) becomes relevant within the context of the so-called WAdS/CFT correspondence, namely the attempt to extend the AdS/CFT correspondence to the case of stretched deformations of AdS$_3$ spaces \cite{Aninnos, DHH}. The reason is that geometries (\ref{sol1})-(\ref{sol2}) permit to explore to what extend the asymptotically WAdS$_3$ boundary conditions can be modified while keeping the conformal symmetry at the boundary unaltered. 

It turns out that the asymptotic behavior of these geometries is substantially different depending on whether $\delta <1$ or $\delta >1$. The relevant components of the metric to see this change of behavior are $g_{rr}$, $g_{r\phi }$, and $g_{\phi \phi }$. For $\delta >1$, these components behave at large $r$ as follows
\begin{eqnarray}
g_{rr} &\simeq & \frac{\ell^2}{(\nu^2+3)r^2} + {\mathcal O}(r^{\delta -4})+ ... \label{jkl1}\\
g_{r\phi } &\simeq & 2\nu r +  {\mathcal O}(r^{\delta -2})+ ... \label{jkl2}\\
g_{\phi \phi } &\simeq & \frac{3}{4}(\nu^2-1) r^2 + {\mathcal O}(r^{\delta })+ ... \label{jkl3} 
\end{eqnarray}
where ${\mathcal O}(r^n)$ stands for terms that damp off as $\sim r^n$ or faster at large $r$, and the ellipses stand for subleading terms. 

Boundary conditions above comprehend a falling-off behavior at large $r$ that is weaker than the one originally proposed in \cite{Compere3}. Nevertheless, as first shown in \cite{Enoc} for the case of TMG, the asymptotic charge algebra associated to the Killing vectors preserving (\ref{jkl1})-(\ref{jkl3}) still coincides with one copy of Virasoro algebra with non-vanishing central charge in semi-direct sum of a current algebra $\hat{u}(1)_k$. In \cite{Proceeding} this analysis was extended to the case of NMG, and it was shown therein how this is consistent with the microscopic counting of WAdS$_3$ black hole microstates in terms of the WAdS$_3$/CFT$_2$ correspondence. 

In contrast, for $\delta <1$, the asymptotic behavior of the metric components is
\begin{eqnarray}
g_{rr} &\simeq & \frac{\ell^2}{(\nu^2+3)r^2} + {\mathcal O}(r^{-3}) + ... \label{FGH1}\\
g_{r\phi } &\simeq & 2\nu r + {\mathcal O}(1) +  ... \label{FGH2}\\
g_{\phi \phi } &\simeq & \frac{3}{4}(\nu^2-1) r^2 + {\mathcal O}(r) +  ... \ , \label{FGH3}
\end{eqnarray}
which is exactly the type of asymptotic considered in \cite{Compere3}. Remarkably, both sets of boundary conditions happen to yield the same asymptotic algebra with the same values of the central charges. While for TMG the Virasoro central charge is given by $c_{\text{TMG}}=(5\nu^2+3)\ell /((\nu^2+3)G\nu ) $, for the case of NMG it is $c_{\text{NMG}}=96\nu^3\ell /((20\nu^4+57\nu^2-9)G)$. This implies that the Cardy formula of the conjectured boundary CFT$_2$ correctly reproduces the entropy of WAdS$_3$ black holes regardless which set of boundary conditions above is considered. This has been recently emphasized in \cite{Proceeding}, and represents a consistency check of the WAdS/CFT correspondence in presence of bulk gravitons -- namely, in presence of asymptotically WAdS$_3$ solutions that, such as (\ref{sol1})-(\ref{sol2}), are not locally equivalent to empty WAdS$_3$ space--. About the latter, notice that having proven here that the time-dependent solutions (\ref{sol1})-(\ref{sol2}) exist in a theory other than TMG was crucial to conclude that a solution with $\omega \neq 0$ is not locally equivalent to WAdS$_3$. Since the curvature invariants of the deformed solution are constant and independent of $H(t,r,\phi ) $, showing that such a solution is associated to the local degrees of freedom of the theory is actually difficult. Here, we managed to prove this by showing that the equations (\ref{polinomio2}) and (\ref{polinomio3}), corresponding to TMG and NMG respectively, yield different conditions for $\omega $ as a function of $\nu $ and $\ell $.

\begin{equation*}
\end{equation*}%
The authors are grateful to Laura Donnay, Andr\'es Goya, Julio Oliva, and Minas Tsoukalas for interesting discussions and collaborations in related subjects. The work of G.G. is partially funded by FNRS-Belgium (convention FRFC PDR T.1025.14 and
convention IISN 4.4503.15), by CONICET, by the Communaut\'{e} Fran\c{c}aise de Belgique through the ARC program and by a donation from the Solvay family. G.G. also thanks Pontificia Universidad Cat\'{o}lica de Valpara\'{\i}so and Universit\'e Libre de Bruxelles, for the hospitality.


\begin{thebibliography}{99}

\bibitem{Proceeding} L. Donnay and G. Giribet, [arXiv:1511.02144 [hep-th]]. 

\bibitem{Enoc} M.~Henneaux, C.~Mart\'{\i}nez and R.~Troncoso, {Phys. Rev. D} {\bf 84} (2011) 124016.

\bibitem{DHH} 
 S.~Detournay, T.~Hartman and D.~M.~Hofman, {Phys. Rev. D} {\bf 86} (2012) 124018.

\bibitem{nos2} L. Donnay and G. Giribet, JHEP {\bf 06} (2015) 99.

\bibitem{BHs} K. Ait Moussa, G. Cl\'{e}ment and C. Leygnac, {Class. Quant. Grav.} {\bf 20} (2003) L277.

\bibitem{TMG} S. Deser, R. Jackiw and S. Templeton, Phys. Rev. Lett. \textbf{%
48} (1982) 975; Ann. Phys. \textbf{140} (1982)\ 372.

\bibitem{NMG} E. Bergshoeff, O. Hohm and P. Townsend, Phys. Rev. Lett. 
\textbf{102} (2009) 201301.

\bibitem{MMG} E. Bergshoeff, O. Hohm, W. Merbis, A. Routh and P. Townsend,
Class. Quant. Grav. \textbf{31} (2014) 145008.

\bibitem{Kuko7} M. Setare, Nucl. Phys. B {\bf 898} (2015) 259.

\bibitem{MMG2} A. Arvanitakis and P. Townsend, Class. Quant. Grav. {\bf 32} (2015) 085003.

\bibitem{otro1} A. Baykal, Class. Quant. Grav. {\bf 32} (2015) 025013 .

\bibitem{matter} A. Arvanitakis, A. Routh and P. Townsend, Class. Quant. Grav. {\bf 31} (2014) 235012.

\bibitem{otro3} M. Alishahiha, M. Qaemmaqami, A. Naseh and A. Shirzad, JHEP {\bf 1412} (2014) 033.

\bibitem{Tekin:2014jna} B.~Tekin, Phys.\ Rev.\ D \textbf{90} (2014) 081701.

\bibitem{Kuko0} G. Giribet and Y. V\'asquez, Phys. Rev. D {\bf 91} (2015) 024026.

\bibitem{Kuko1} M. Setare and H. Adami, [arXiv:1507.00107 [hep-th]].

\bibitem{Kuko2} N. Deger and O. Sarioglu, [arXiv:1505.03387 [hep-th]].

\bibitem{Kuko3} M. Setare and H. Adami, Phys. Lett. B {\bf 744} (2015) 280.

\bibitem{Kuko4} E. Altas and B. Tekin, Phys. Rev. D {\bf 92} (2015) 025033.

\bibitem{Kuko5} A. Arvanitakis, Class. Quant. Grav. {\bf 32} (2015) 115010.

\bibitem{Kuko6} M. Setare and H. Adami, Phys. Rev. D {\bf 91} (2015) 104039.

\bibitem{Kuko8} B. Tekin, Phys. Rev. D {\bf 92} (2015) 024008.

\bibitem{Aninnos} D.~Anninos, W.~Li, M.~Padi, W.~Song and A.~Strominger, {JHEP} {\bf 0903} (2009) 130.

\bibitem{Clement} G.~Cl\'ement, {Class. Quant. Grav.}  {\bf 26} (2009) 105015.

\bibitem{Eloy} A. Ay\'{o}n-Beato and M. Hassa\"{\i}ne, Annals Phys. \textbf{317} (2005) 175; Phys. Rev. D \textbf{73} (2006) 104001. 

\bibitem{AdSwaves} A. Ay\'{o}n-Beato,
G. Giribet and M. Hassa\"{\i}ne, JHEP \textbf{0905} (2009) 029.

\bibitem{GAY} A. Garbarz, G. Giribet and Y. V\'{a}squez, Phys. Rev. D 
\textbf{79} (2009)\ 044036.

\bibitem{Compere3} 
  G.~Comp\`{e}re and S.~Detournay, {JHEP} {\bf 0908} (2009) 092.

\bibitem{ADT1} L. F. Abbott and S. Deser, Nucl. Phys. B \textbf{195} (1982)
76.

\bibitem{ADT2} S. Deser and B. Tekin, Phys. Rev. Lett. \textbf{89} (2002)
101101; Phys. Rev. D \textbf{67} (2003) 084009; Class. Quant. Grav. \textbf{%
19} (2002) L97; ibid. \textbf{20} (2003) L259.

\bibitem{Barnich:2001jy} G.~Barnich and F.~Brandt, Nucl.\ Phys.\ B {\bf 633} (2002) 3.

\bibitem{Barnich:2007bf} G.~Barnich and G.~Comp\`{e}re, J.\ Math.\ Phys.\  {\bf 49} (2008) 042901.

\bibitem{IyerWald} V.~Iyer and R.~M.~Wald, Phys.\ Rev.\ D {\bf 50} (1994) 846.

\bibitem{Clemento}  A. Bouchareb and G. Cl\'ement, Class. Quant. Grav.
{\bf 24} (2007) 5581.

\bibitem{Corea} 
 S.~Nam, J.~D.~Park and S.~H.~Yi, {Phys. Rev. D} {\bf 82} (2010) 124049.

\bibitem{Minas} G. Giribet and M. Tsoukalas, Phys. Rev. D {\bf 92} (2015) 064027.



\end{thebibliography}
\end{document}